\documentclass[aps,prb,twocolumn,showpacs,preprintnumbers,amsmath,amssymb,superscriptaddress]{revtex4}%
\usepackage{graphicx}% Include figure files
\usepackage{dcolumn}% Align table columns on decimal point
\usepackage{bm}% bold math
\usepackage{color}
\begin{document}

%\preprint{PREPRINT (\today)}

\title{A tool to estimate the critical dynamics and thickness of
superconducting films and interfaces}

\author{T. Schneider}
\email{tschnei@physik.unizh.ch}
\affiliation{Physik-Institut der Universit\"{a}t Z\"{u}rich, Winterthurerstrasse 190, CH-8057 Z\"{u}rich, Switzerland}

\begin{abstract}
We demonstrate that the magnetic field dependence of the conductivity
measured at the transition temperature allows the dynamical critical
exponent, the thickness of thin superconducting films and interfaces, and
the limiting lateral length to be determined. The resulting tool is applied
to the conductivity data of an amorphous Nb$_{0.15}$Si$_{0.85}$ film and a LaAlO$_{3}$/SrTiO$_{3}$ interface.
\end{abstract}

\pacs{74.78.-w, 74.40.+k, 74.90.+n, 74.78.Fk}
\maketitle

In a phase transition, sufficiently close to the transition temperature $T_{c}$, critical fluctuations are expected to dominate. The closer one gets to $T_{c}$, the longer these fluctuations will last, and the larger the
relevant length scale becomes. In a superconductor the relevant length scale
is the correlation length $\xi $. Without loss of generality we can assume
that the lifetime of the fluctuations, $\tau $, varies as $\tau \propto \xi
^{z}$ which defines $z$, the dynamical critical exponent.\cite{ffh,book} As
we approach the critical region, all the physics that really matters is
associated with the diverging length and time scales.

Using experimentally accessible quantities, voltage $V$ and current $I$,
dynamic scaling predicts for superconducting films and interfaces the
relationship\cite{ffh}
\begin{equation}
V=I\xi ^{-z}g_{\pm }\left( \frac{I\xi }{T}\right) .  \label{eq1}
\end{equation}
$g_{\pm }\left( x\right) $ is a scaling function of its argument above (+)
and below (-) $T_{c}$. Above $T_{c}$, in the limit $x\rightarrow 0$, $g_{+}\left( x\right) $ tends to a constant and the conductivity to
\begin{equation}
\sigma =\frac{I}{V}\propto \xi ^{z}.  \label{eq2}
\end{equation}%
On the other hand, at $T_{c}$ in the limit $x\rightarrow \infty $, $g_{\pm}\left( x\right) $ tends to $x^{z}$ so that
\begin{equation}
V\propto I^{a\left( T_{c}\right) },\text{ }a\left( T_{c}\right) =z+1.
\label{eq3}
\end{equation}
In practice $I-V$-data exhibit resistive tails revealing finite size induced
free vortices which make it difficult to estimate the transition temperature
$T_{c}$ and the dynamical scaling exponent $z$.\cite{pierson,med,holzer,strachan,hao}

Alternatively, the application of the conductivity relation (\ref{eq2})
requires the explicit form of the correlation length. Since superconducting thin films and interfaces are expected to undergo a Berezinskii-Kosterlitz-Thouless (BKT) transition from the superconducting to the normal state the correlation length adopts for $T\geq T_{c}$ the characteristic form\cite{bere,kosterlitz}
\begin{equation}
\xi \left( T\right) =\xi _{0}\exp \left( \frac{2\pi }{bt^{1/2}}\right) ,t=\frac{T}{T_{c}}-1.
\label{eq4}
\end{equation}
$\xi _{0}$ is related to the vortex core radius and $b$ to the energy needed
to create a vortex.\cite{ambeok,finotello,steele,tilly} Accordingly the
analysis of conductivity or resistivity data in zero magnetic field provide in terms of $\sigma \propto \xi ^{z}$ estimates for $T_{c}$, $\xi_{0}^{z}$ and $z/b$ \cite{science,nature,apl,tsprb}, while the dynamical
critical exponent $z$ cannot be determined. Furthermore, the relationship $\sigma \propto \xi ^{z}$ allows to perform a standard finite size scaling analysis.\cite{tsprb,privman}

In this context it is important to recognize that the existence of the
BKT-transition (vortex-antivortex dissociation instability) in $^{4}$He films
is intimately connected with the fact that the interaction
energy between vortex pairs depends logarithmic on the separation between
them. As shown by Pearl\cite{pearl}, vortex pairs in thin superconducting
films (charged superfluid) have a logarithmic interaction energy out to the
characteristic length $\lambda _{2D}=\lambda ^{2}/d$, beyond which the
interaction energy falls off as $1/r$. Here $\lambda $ is the magnetic
penetration depth of the bulk. As $\lambda _{2D}$ increases the diamagnetism of the superconductor
becomes less important and the vortices in a thin superconducting film
become progressively like those in $^{4}$He films.\cite{beasley} According to
this $\lambda _{2D}>>\min \left[ W,\text{{}}L\right] $ is required, where $W$
and $L$ denote the width and the length of the perfect sample. Invoking the
Nelson-Kosterlitz relation\cite{nelson} $\lambda _{2D}\left( T_{c}\right)
=\lambda ^{2}\left( T_{c}\right) /d=\Phi _{0}^{2}/\left( 32\pi
^{2}k_{B}T_{c}\right) $ it is readily seen that for sufficiently low $T_{c}$%
's and $\min \left[ W,\text{{}}L\right] <<1$ cm this condition is well
satisfied. As a result any rounding of the transition due to finite size
effects should be more important than that due to the finite magnetic
"screening length" $\lambda _{2D}$.

Here we present a tool to determine the dynamical critical exponent $z$, the
thickness $d$, and the limiting length $\widehat{L}$, associated with the
resistive tail in zero magnetic field, from conductivity measurements taken
at $T_{c}$ and in magnetic fields applied parallel and perpendicular to the
film or interface. Traditionally the thickness of superconducting films is
estimated from the angular dependence of the upper critical field $H_{c2}$.\cite{tinkham} Noting that $H_{c2}$ is an artifact of the mean-field approximation this approach becomes questionable in two dimensions where
thermal fluctuations are enhanced. The crucial component of the tool stems
from the magnetic field induced finite size effect. For $T\geq T_{c}$ and
nonzero magnetic field the mean distance between the vortex lines $\left(
\Phi _{0}/H\right) ^{1/2}$ is another characteristic length, preventing the
correlation length to diverge at $T_{c}$ and $H>0$.\cite{tsjpc} The
resulting magnetic field induced finite size effect can be described by
relating the zero field and finite field correlation length in terms of
\begin{equation}
\xi _{x}\left( T,H_{z}\right) \xi _{y}\left( T,H_{z}\right) =\xi _{x}\left(
T,0\right) \xi _{y}\left( T,0\right) G\left( x\right) ,
\label{eq5}
\end{equation}
where
\begin{eqnarray}
x &=&\frac{aH_{z}\xi _{x}\left( T,0\right) \xi _{y}\left( T,0\right) }{\Phi
_{0}}=\frac{\xi _{x}\left( T,0\right) \xi _{y}\left( T,0\right) }{%
L_{H_{z}}^{2}},  \nonumber \\
L_{H_{z}}^{2} &=&\frac{\Phi _{0}}{aH_{z}}.
\label{eq6}
\end{eqnarray}
$L_{H_{z}}$ is the limiting magnetic length and $G\left( x\right) $ denotes the
finite size scaling function with the limiting behavior
\begin{equation}
G\left( x\right) =\left\{
\begin{array}{c}
1\text{ \ }:x=0\text{ } \\
1/x:x\rightarrow \infty
\end{array}
\right. .  \label{eq7}
\end{equation}
Indeed, in zero field the limiting magnetic length $L_{H_{z}}$ is infinite
and the growth of the correlation length $\xi $ is unlimited, while in
finite fields the divergence of $\xi $ at $T_{c}$is removed and its value is
given by
\begin{equation}
\xi _{x}\left( T_{c},H_{z}\right) \xi _{y}\left( T_{c},H_{z}\right)
=L_{H_{z}}^{2}=\frac{\Phi _{0}}{aH_{z}},
\label{eq8}
\end{equation}
where $a$ fixes the mean distance between vortices. The equivalence to the
standard finite size effect in a film of dimensions $L\times L$ is readily
established by noting that in this case the correlation length scales as $\xi \left( T,L\right) =\xi \left( T,L=\infty \right) G\left( \xi \left(T,L=\infty \right) /L\right) $.\cite{privman}

More generally in magnetic fields $H_{\bot ,\Vert }$, applied perpendicular ($\bot $) or parallel ($\Vert $) to the film or interface, the divergence of $\xi \left( T\right) $ at $T_{c}$ is then removed because $\xi \left(
T_{c}\right) $ cannot grow beyond
\begin{equation}
\widetilde{L}=\left\{
\begin{array}{ccc}
\widehat{L} &  &  \\
L_{H_{\bot }} & = & \left( \frac{\Phi _{0}}{aH_{\bot }}\right) ^{1/2} \\
L_{H_{\Vert }} & = & L_{H_{\Vert }}=\frac{\Phi _{0}}{aH_{\Vert }d}
\end{array}
\right. .
\label{eq9}
\end{equation}
Here we included the limiting length $\widehat{L}$ arising from the ohmic
tail in zero field, e.g. due to the system size or the finite lateral extent
of the homogenous domains. The expressions for the magnetic field induced
limiting lengths $L_{H_{\bot }}$ and $L_{H_{\Vert }}$ follow from Eq. (\ref{eq8}) and by noting that the correlation lengths of fluctuations which are transverse to the applied magnetic field are bounded according to $\xi
_{x}\xi _{y}\leq \Phi _{0}/\left( aH_{x}\right) $, $x\neq y\neq z$, where $\xi _{z}=d$, $H_{\bot }=H_{z}$, $H_{x}=H_{y}=H_{\Vert }$, and accordingly $\xi _{x}\xi _{y}=\xi _{\Vert }^{2}\leq L_{H_{\bot }}^{2}=\Phi _{0}/aH_{\bot
} $ and $\xi _{x}\xi _{z}=\xi _{\Vert }d\leq L_{H_{\Vert }}d=\Phi
_{0}/aH_{\Vert }$, where $d$ denotes the film thickness.

These limiting lengths prevent the divergence of the conductivity at $T_{c}$. In zero field it adopts according to Eqs. (\ref{eq2}) and (\ref{eq9}) the
form%
\begin{equation}
\sigma \left( T_{c},H_{\bot ,\Vert }=0\right) =f\text{ }\widehat{L}^{z},
\label{eq10}
\end{equation}
As the magnetic field increases this behavior applies as long as $\widehat{L}<$ $L_{H_{\bot ,\Vert }}$, while for $\widehat{L}>$ $L_{H_{\bot ,\Vert }}$
the magnetic field sets the limiting length and the conductivity approaches
according to Eqs. (\ref{eq2}) and (\ref{eq9}) the form
\begin{equation}
\sigma \left( T_{c},H_{\bot ,\Vert }\right) =\sigma _{n}+\left\{
\begin{array}{c}
f_{\bot }H_{\bot }^{-z/2},f_{\bot }=f\left( \Phi _{0}/a\right) ^{z/2} \\
f_{\Vert }H_{\Vert }^{-z}\text{ \ },f_{\Vert }=f\left( \Phi _{0}/ad\right)
^{z}%
\end{array}
\right. ,
\label{eq11}
\end{equation}
where $\sigma _{n}$ is the normal state conductivity, attained in the high
field limit. The thickness $d$ of the superconducting film or interface
follows then from%
\begin{equation}
d^{2}=\frac{\Phi _{0}}{a}\left( \frac{f_{\bot }}{f_{\Vert }}\right) ^{2/z},
\label{eq12}
\end{equation}
whereby an estimation of $d$ requires the value of the dynamical critical
exponent $z$, derivable from the magnetic field dependence of the
conductivity at $T_{c}$ (Eq. (\ref{eq11})). So far we concentrated on
temperatures at and above the BKT-transition. Below $T_{c}$ the correlation
length diverges: $\xi \rightarrow \infty $.\cite{bere,kosterlitz} This implies that
$\xi $ will be cut off by a limiting length and with that are Eqs. (\ref{eq10}) and (\ref{eq11}) expected to apply for $0<T\leq T_{c}$. Since the low-temperature phase in the BKT scenario is described by a line of fixed points, each
temperature $T<T_{c}$ may be characterized by its own $f\left( T\right) $.

An essential assumption of the outlined approach is that around $T_{c}$
thermal phase fluctuations dominate. There is considerable evidence for a
critical magnetic field $H_{_{\bot ,\Vert }c}$, emerging from a nearly
temperature independent crossing point in the resistance-magnetic field
plane.\cite{kim,marko,gant,mason,biele,gold,aubin,steiner} It can be identified as the critical field
of the quantum superconductor to insulator (QSI) transition and the
resistance is predicted to scale as $R\left( H_{\bot ,\Vert },T\right)
=R_{c}f\left( \left\vert H_{\bot ,\Vert }-H_{_{\bot ,\Vert }c}\right\vert
/T^{1/\overline{z}\overline{\nu }}\right) $,\cite{sondhi} where $\overline{\nu }$ is the zero temperature correlation length exponent and $\overline{z}$ the quantum dynamical critical exponent. However, recent experiments\cite{aubin,crane} that have explored the competition between thermal and quantum fluctuations at low enough temperatures revealed that a temperature independent critical field occurs at low temperatures only, where quantum fluctuations are no longer negligible.
\begin{figure}[htb]
%\centering
\includegraphics[width=1.0\linewidth]{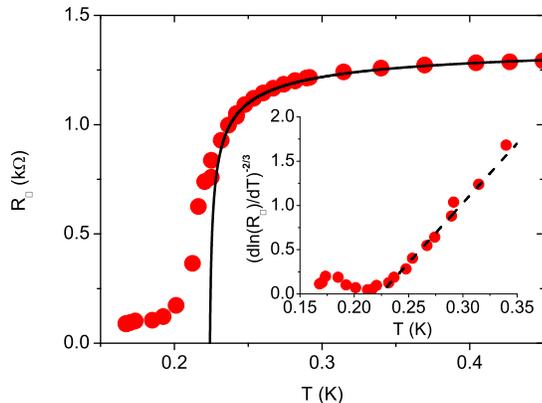}
\vspace{-1.0cm}
\caption{$R_{\square }\left( T\right) $ of an amorphous $125$ \AA\ thick Nb$_{0.15}$Si$_{0.85}$ film taken from Aubin \textit{et al}. \cite{aubin} The solid line is $R=R_{0}\exp \left( -b_{R}/\left( T-T_{c}\right) ^{1/2}\right)
$ with $R_{0}=1.41$ k$\Omega $, $b_{R}=0.0403$ K$^{1/2}$, and $T_{c}=0.224$ K. The inset shows $(d\ln \left( R\right) /dT)^{-2/3}$ \textit{vs}. $T$ and the dashed line is $(d\ln \left( R\right) /dT)^{-2/3}=\left( 2/b_{R}\right)
^{2/3}\left( T-T_{c}\right) $.}
\label{fig1}
\end{figure}

To illustrate this tool, allowing $z$, $d$, and $\widehat{L}$ to be
determined from the magnetic field dependence of the conductivity at $T_{c}$
we analyze next the data of Aubin \textit{et al}. \cite{aubin} of an
amorphous $125$ \AA\ thick Nb$_{0.15}$Si$_{0.85}$ film. In Fig.\ref{fig1} we
depicted the temperature dependence of the sheet resistance in zero field to
estimate $T_{c}$ and to uncover a rounded transition attributable to a
finite size effect. Evidence for characteristic BKT-behavior emerges from
the inset showing $(d\ln \left( R\right) /dT)^{-2/3}$ vs. $T$ in terms of
the consistency with $(d\ln \left( R\right) /dT)^{-2/3}=\left(
2/b_{R}\right) ^{2/3}\left( T-T_{c}\right) $ in an intermediate temperature
regime above $T_{c}$. The resulting estimates for $b_{R}$ and $T_{c}$ are
then used to obtain the BKT-resistance, $R=R_{0}\exp \left( -b_{R}/\left(
T-T_{c}\right) ^{1/2}\right) $, by adjusting $R_{0}$ in this intermediate
regime. The comparison between the resulting solid BKT-line and the data
reveals a rounded transition and with that a finite size effect
generating free vortices at and below $T_{c}=0.224$ K. In this context we note that
according to the Harris criterion weak randomness in the local $T_{c}$,
pairing interaction, etc. does not change the critical BKT-behavior.\cite{aharony} Nevertheless, inhomogeneities due to local strain or a heat current appear to be likely in both, superconducting films and interfaces. A
nonzero heat current drives the system away from equilibrium. A temperature
gradient is created which implies that the temperature is space dependent.

\begin{figure}[htb]
%\centering
\includegraphics[width=1.0\linewidth]{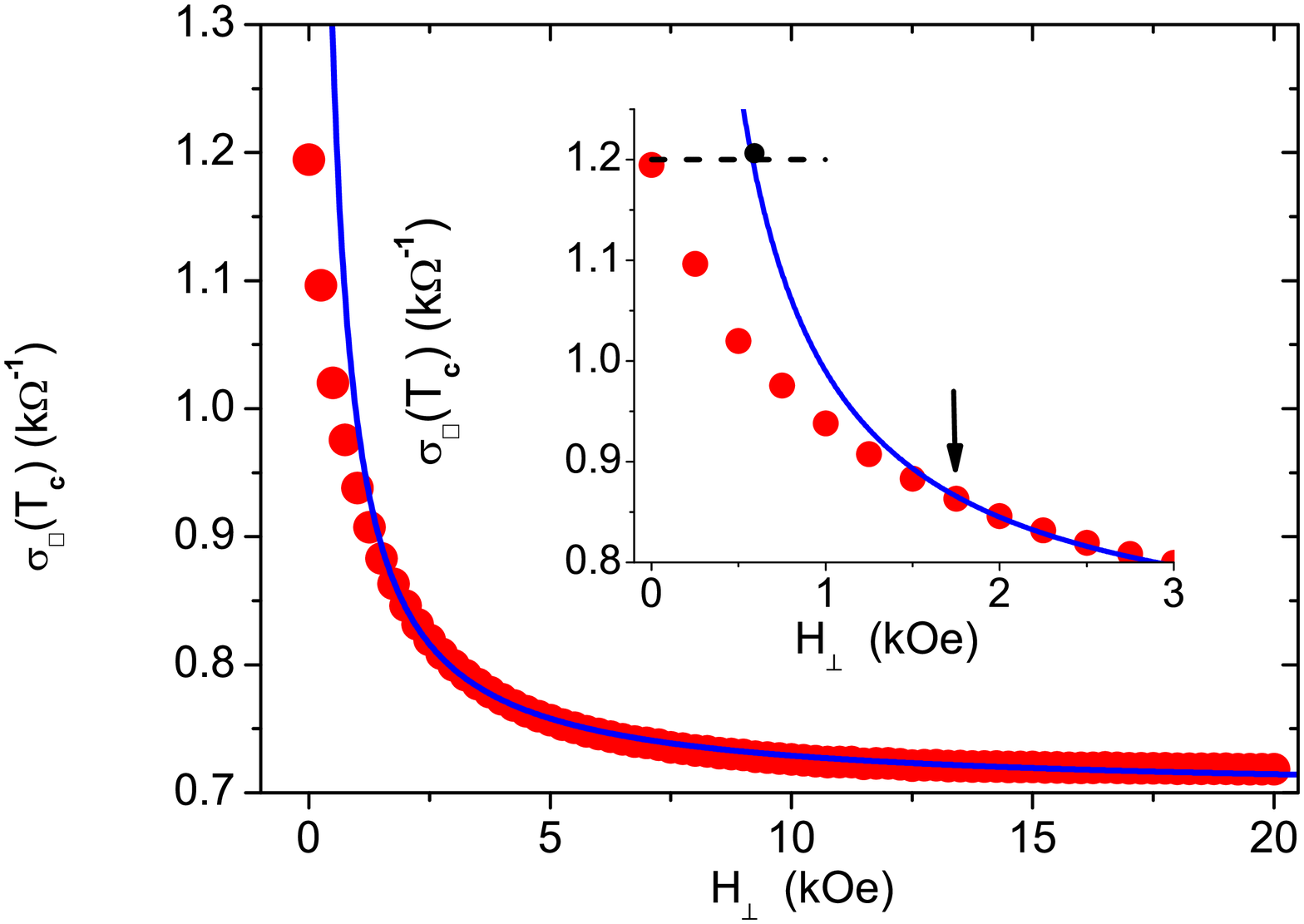}
\vspace{-1.0cm}
\caption{$\sigma _{\square }\left( T_{c}\right) $ \textit{vs} $H_{\bot }$ for an amorphous $125$ \AA\ thick Nb$_{0.15}$Si$_{0.85}$ film and $T\simeq 0.224$ K$\simeq T_{c}$ derived from Aubin \textit{et al}. \cite{aubin} The solid line is Eq. (\ref{eq11}) with $\sigma _{n}=0.70$ k$\Omega ^{-1}$ and $f_{\bot }=0.29$ k$\Omega $$^{-1}$. The arrow marks $H_{\bot}^{\ast }=1.75$ kOe and the dot $H_{\bot }^{\bullet }=0.59 $kOe.}
\label{fig2}
\end{figure}

 Given the estimate of the BKT-transition temperature $T_{c}$ and the
evidence for a zero field limiting length we turn to the effects of an
applied magnetic field, inducing additional free vortices. In Fig.\ref{fig2} we show
the sheet conductivity $\sigma _{\square }\left( T_{c}\right) $ \textit{vs}.
$H_{\bot }$ derived from the resistivity data. Above $H_{\bot }^{\ast }=1.75$ kOe we observe for
\begin{equation}
z\simeq 2,
\label{eq13}
\end{equation}
consistency with $\sigma \left( T_{c},H_{\bot }\right) =\sigma_{n}+f_{\bot }H_{\bot }^{-z/2}$ (Eq. (\ref{eq11})) and therewith evidence for diffusive dynamics.\cite{ffh} In the low field limit deviations from Eq. (\ref{eq11}) are expected because for sufficiently low $H_{\bot }$ the magnetic length $L_{H_{\bot }}=\left( \Phi _{0}/aH_{\bot }\right) ^{1/2}$ is no longer large compared to $\widehat{L}$, the zero field limiting length.

\begin{figure}[htb]
%\centering
\includegraphics[width=1.0\linewidth]{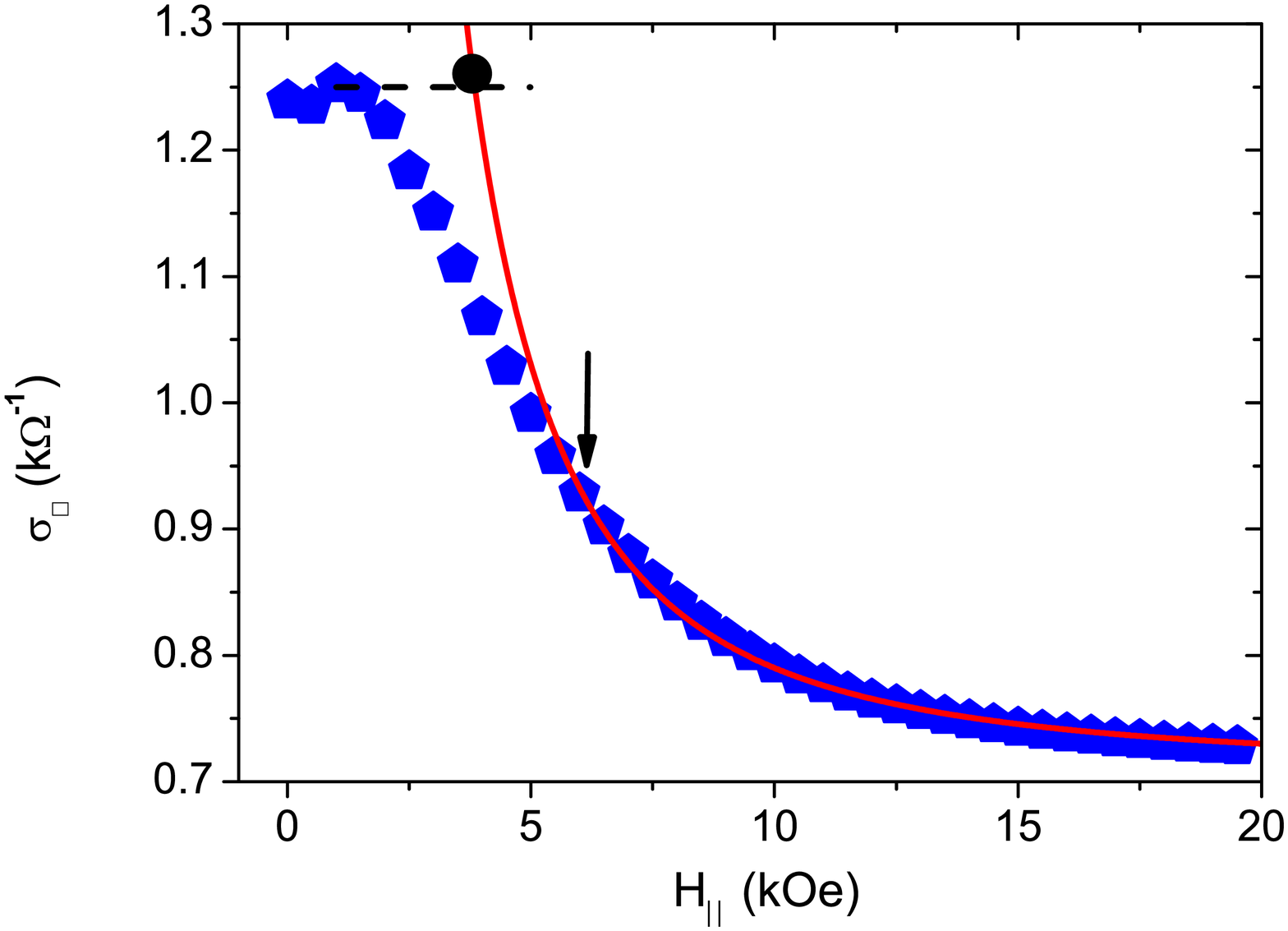}
\vspace{-1.0cm}
\caption{$\sigma _{\square }\left( T_{c}\right) $ \textit{vs}. $%
H_{\Vert }$ for an amorphous $125$ \AA\ thick Nb$_{0.15}$Si$_{0.85}$ film
and $T=0.224$ K$\simeq T_{c}$ derived from Aubin \textit{et al}.\cite{aubin}
The solid line is Eq. (\ref{eq11}) with $\sigma _{n}=0.71$ k$\Omega ^{-1}$
and $f_{\Vert }=8$ k$\Omega ^{-1}$kOe$^{2}$. The arrow marks $H_{\Vert
}^{\ast }=6$ kOe and the dot $H_{\Vert }^{\bullet }=3.85$ kOe.}
\label{fig3}
\end{figure}

According to Fig.\ref{fig3}, depicting $\sigma _{\square }\left( T_{c}\right) $
\textit{vs}. $H_{\Vert }$ of the same sample, agreement with $\sigma \left(T_{c},H_{\Vert }\right) =\sigma _{res}+f_{\Vert }H_{\Vert }^{-z}$ (Eq. (\ref{eq11})) is obtained above $H_{\Vert }^{\ast }=6$ kOe for $z\simeq 2$. So
this value is consistent with both the perpendicular and parallel magnetic
field dependence. Given then the evidence for $z=2$ and the estimates for $f_{\bot }$ and $f_{\Vert }$ we obtain with the nominal thickness of the film, $d\simeq 125$\AA\ \cite{aubin} and Eq. (\ref{eq12}) for $a$, fixing
the mean distance between vortices, the estimate
\begin{equation}
a\simeq 4.8,  \label{eq14}
\end{equation}
compared to $a\simeq 3.12$, found in bulk cuprate superconductors.\cite{tsjpc} Note that the film thickness was monitored in situ during the evaporation
by a set of piezoelectric quartz. Moreover, the thicknesses and compositions
were checked ex situ by Rutherford backscattering. The accuracy is estimated
to be $\pm $ 5\%.\cite{marra} In analogy to the behavior in the
perpendicular field deviations from Eq. (\ref{eq11}) occur with reduced
field strength. They set in around $H_{\Vert }^{\ast }=6$ kOe where $L_{H_{\Vert }^{\ast }}=\Phi _{0}/\left( adH_{\Vert }^{\ast }\right) $ is no longer large compared to $\widehat{L}$. To estimate $\widehat{L}$ we note
that Eqs. (\ref{eq10}) and (\ref{eq11}) imply that at $H_{\Vert }^{\bullet }$
and $H_{\bot }^{\bullet }$ (see Figs. \ref{fig2} and \ref{fig3}) the relation
\begin{equation}
\widehat{L}=\frac{\Phi _{0}}{adH_{\Vert }^{\bullet }}=\left( \frac{\Phi _{0}}{aH_{\bot }^{\bullet }}\right) ^{1/2}
\label{eq15}
\end{equation}
holds. With $H_{\bot }^{\bullet }=0.59$kOe and $a\simeq 4.8$ we obtain $\widehat{L}\simeq 855$\AA , while $H_{\Vert }^{\bullet }=3.85$ kOe and $d=125$ \AA\ yields $\widehat{L}\simeq 896$\AA , compared to the lateral
dimensions $W\times L=0.28$ cm$\times 0.15$ cm of the film.\cite{marra} Invoking the
Kosterlitz-Nelson relation $\lambda _{2D}\left( T_{c}\right) =\lambda
^{2}\left( T_{c}\right) /d=\Phi _{0}^{2}/\left( 32\pi ^{2}k_{B}T_{c}\right) $
we obtain $\lambda _{2D}\left( T_{c}\right) \simeq 4.4$ cm for $T_{c}=0.224$
K, whereupon $\lambda _{2D}>>\min \left[ W,\text{{}}L\right] $ is well
satisfied for this film. Because $\lambda _{2D}\left(
T_{c}\right) $ is also large compared to $\widehat{L}$, the zero field
limiting length appears to be set by the lateral extent of the homogenous
domains. In any case, the uncovered limiting length implies the presence of free vortices below $T_{c}$, precluding a true phase transition. Accordingly, the rounded BKT-transition seen in Fig. \ref{fig1} is traced back to a limiting
length not attributable to the finite magnetic "screening length" $\lambda _{2D}$.

\begin{figure}[htb]
%\centering
\includegraphics[width=1.0\linewidth]{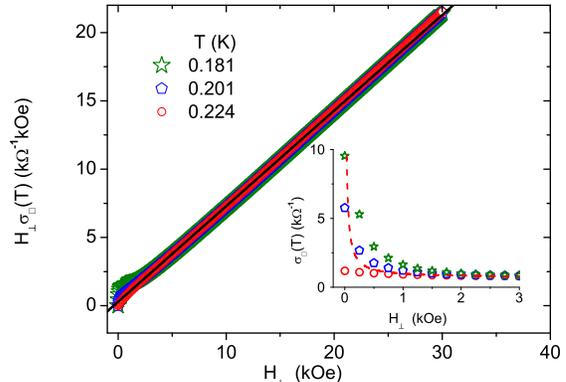}
\vspace{-1.0cm}
\caption{$H_{\bot }\sigma _{\square }\left( T\right) $ \textit{vs}.
$H_{\bot }$ for an amorphous $125$ \AA\ thick Nb$_{0.15}$Si$_{0.85}$ film at
$T=0.224$ K$\simeq T_{c}$ , $T=0.201$ K, and $T=0.181$ K derived from Aubin
\textit{et al}.\cite{aubin} The solid line is Eq. (\ref{eq11}) in terms of $H_{\bot }\sigma \left( T_{c},H_{\bot }\right) =\sigma _{n}H_{\bot }+f_{\bot }$ with $z=2$, $\sigma _{n}=0.70$ k$\Omega ^{-1}$ and $f_{\bot }=0.29$ k$%
\Omega $. The inset shows $\sigma _{\square }\left( T\right) $ \textit{vs}. $H_{\bot }$. The dashed line is Eq. (\ref{eq11}) with $\sigma _{n}=0.70$ k$\Omega ^{-1}$ and $f_{\bot }=0.29$ k$\Omega ^{-1}$kOe}
\label{fig4}
\end{figure}

As aforementioned, below $T_{c}$ the correlation length diverges.\cite{bere,kosterlitz} Correspondingly, $\xi \rightarrow \infty $ will be cut off by a limiting length and Eqs. (\ref{eq10}) and (\ref{eq11}) are expected to
apply for $T\leq T_{c}$. Since the low-temperature phase in the BKT scenario
is described by a line of fixed points, each temperature $T<T_{c}$ may be
characterized by its own $f\left( T\right) $. To clarify this conjecture we
invoke Eq. (\ref{eq11}) in the form $H_{\bot }\sigma \left( T,H_{\bot
}\right) =H_{\bot }\sigma _{n}+f_{\bot }\left( T\right) $ with $z=2$. The
data should then fall on straight lines with slope $\sigma _{n}$ and
intercepts $f_{\bot }\left( T\right) $. In Fig. \ref{fig4}, depicting $H_{\bot
}\sigma _{\square }\left( T\right) $ \textit{vs}. $H_{\bot }$ for
temperatures at and below $T_{c}$, we observe that above $H_{\bot }^{\ast
}=1.75$ kOe (see Fig. \ref{fig2}), where the magnetic field sets the limiting length,
the data falls on a single line, while below $H_{\bot }^{\ast }$ a crossover
to the zero field limit behavior, $\sigma \left( T_{c},H_{\bot }=0\right)
=f\left( T\right) \widehat{L}^{z}$ (Eq. (\ref{eq10})) sets in. Indeed,
around $H_{\bot }^{\ast }$ the magnetic limiting length $L_{H_{\bot }}$
becomes comparable to $\widehat{L}$. From the inset, showing $\sigma
_{\square }\left( T\right) $ \textit{vs}. $H_{\bot }$, it is seen that in
zero field $f\left( T\right) $ increases with reduced temperature,
reflecting that by lowering the temperature the density of the finite size
induced vortices is reduced and with that the conductivity increases. Thus,
as conjectured, $f\left( T\right) $ in Eq. (\ref{eq10}) depends on
temperature. The agreement with Eq. (\ref{eq11}), taking thermal
fluctuations into account only, also reveals that around $T_{c}$ the
contribution of quantum fluctuations is negligibly small, although a nearly
temperature independent crossing point in the resistance-magnetic field
plane occurs around $H_{\bot }\simeq 5.5$kOe.\cite{aubin}

  To illustrate this tool further, allowing to determine $z$, $d$ and $\widehat{L}$ from the magnetic field dependence of the conductivity at $T_{c} $ we analyze the conductivity data of Reyren \textit{et al}. \cite{apl}
for a superconducting LaAlO$_{3}$/SrTiO$_{3}$\ interface \ with $T_{c}\simeq0.21 $ K. In Fig. 5 we show the sheet conductivity $\sigma _{\square }\left(T_{c}\right) $ \textit{vs}. $H_{\bot }$ derived from the resistivity data.
Above $\mu _{0}H_{\bot }\simeq 10$ mT we observe consistency with Eq. (\ref{eq11}) for $z\simeq 2$, in agreement with the value derived from I-V-data, \cite{science} and predicted for diffusive dynamics. \cite{ffh} According to
Fig. 6 and Eq. (\ref{eq11}) $z\simeq 2$ also follows from $\sigma \left(T_{c}\right) $ \textit{vs}. $H_{\Vert }$ above $\mu _{0}H_{\Vert }\gtrsim 300 $ mT. \ Given then the evidence for $z=2$ and the estimates for $f_{\bot}$ and $f_{\Vert }$ we obtain with Eqs. (\ref{eq12}) and (\ref{eq14}) for the thickness of the superconducting interface the value
\begin{equation}
d\simeq 67\text{\AA ,}
\label{eq16}
\end{equation}
in agreement with previous estimates where $z=2$ was assumed.\cite{apl} Recently, room temperature studies have also been performed to estimate the thickness of the LaAlO$_{3}$/SrTiO$_{3}$\ interface grown at
\textquotedblleft high\textquotedblright\ oxygen pressures leading to a value of $70$\AA\ \cite{baslet} , $100$\AA \cite{sing} and $120$\AA\ at 8 K.\cite{copie}
\begin{figure}[htb]
%\centering
\includegraphics[width=1.0\linewidth]{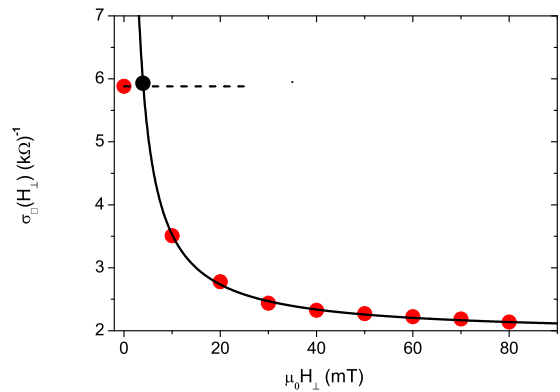}
\vspace{-1.0cm}
\caption{ $\sigma _{\square }\left( T_{c}\right) $ \textit{vs} $H_{\bot }$
for a LaAlO$_{3}$/SrTiO$_{3}$\ interface with $T_{c}\simeq 0.21$ K derived
from Reyren \textit{et al}. \cite{apl} The solid line is Eq. (\ref{eq11})
with $\sigma _{n}=1.94\times 10^{-3}\Omega $ and $f_{\bot }=1.59\times 10^{-2\text{ }}\Omega $mT. The dot marks $\mu _{0}H_{\bot }^{\bullet }=3.8$ mT.}
\label{fig5}
\end{figure}

\begin{figure}[htb]
%\centering
\includegraphics[width=1.0\linewidth]{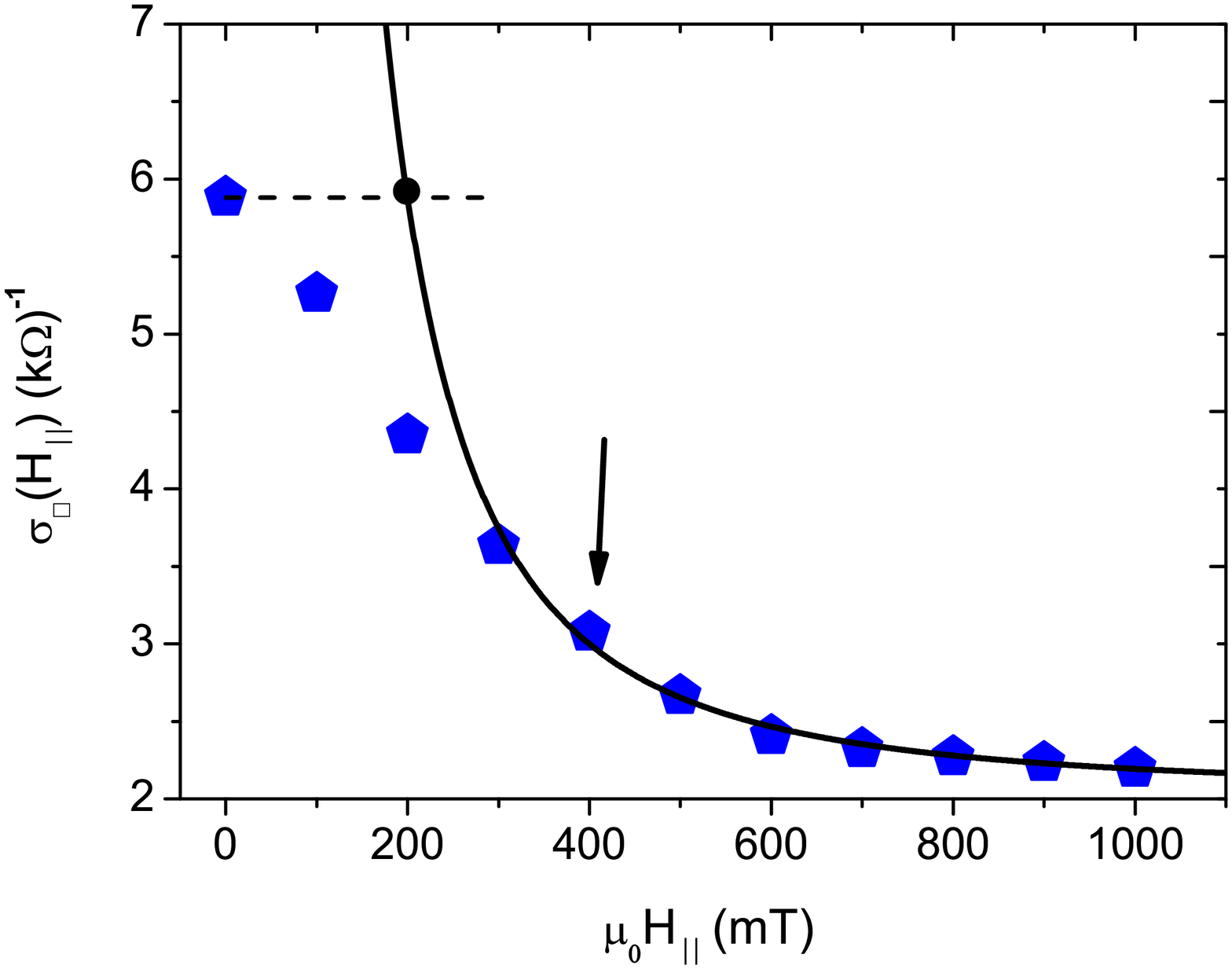}
\vspace{-1.0cm}
\caption{$\sigma \left( T_{c}\right) $ \textit{vs} $H_{\Vert }$ for a LaAlO%
$_{3}$/SrTiO$_{3}$\ interface \ with $T_{c}\simeq 0.21$ K derived from
Reyren \textit{et al}. \cite{apl}. The solid line is Eq. (\ref{eq11}) with $%
\sigma _{n}=2.04\times 10^{-3}$ $\Omega $ and $f_{\bot }=153.42$ $\Omega $mT$%
^{2}$. The dot marks $\mu _{0}H_{\Vert }^{\bullet }=195$ mT.}
\label{fig6}
\end{figure}

Furthermore, in analogy to the amorphous Nb$_{0.15}$Si$_{0.85}$ film (see
Figs. \ref{fig2} and \ref{fig3}) $\sigma _{\square }\left( T_{c}\right) $
\textit{vs}. $H_{\bot ,\Vert }$ does not diverge in the zero field limit.
This behavior was traced back to a standard finite size effect, presumably
attributable to a finite lateral extent $\widehat{L}$ of the homogeneous
domains. \cite{tsprb} To substantiate this interpretation we invoke Eq. (\ref{eq15}) and the respective estimates for $H_{\bot }^{\bullet }$ and $H_{\Vert }^{\bullet }$, yielding with $a=4.8$ and $d\simeq 67$\AA, $\widehat{L}\simeq 3.4\times 10^{-5}$ cm ($\mu _{0}H_{\bot }^{\bullet }=3.8$ mT) and $\widehat{L}\simeq 4.9\times 10^{-5}$ cm ($\mu _{0}H_{\Vert }^{\bullet }=195$mT), compared to the lateral dimensions $W\times L=0.02$ cm$\times 0.01$ cm
of the superconducting interface. Invoking the Kosterlitz-Nelson relation $\lambda _{2D}\left( T_{c}\right) =\lambda ^{2}\left( T_{c}\right) /d=\Phi_{0}^{2}/\left( 32\pi ^{2}k_{B}T_{c}\right) $ we obtain $\lambda _{2D}\left(
T_{c}\right) \simeq 4.8$ cm for $T_{c}=0.21$ K, whereupon $\lambda
_{2D}>>\min \left[ W,\text{{}}L\right] $ is well satisfied for the LaAlO$_{3}
$/SrTiO$_{3}$\ interface. Furthermore, because $\lambda _{2D}\left(
T_{c}\right) $ is also large compared to $\widehat{L}$, the zero field
limiting length appears to be set by the lateral extent of the homogenous
domains. In any case due to the uncovered limiting length, not attributable
a finite magnetic "screening length" $\lambda _{2D}$, it becomes possible
for free vortices to form below $T_{c}$ which in turn precludes a true phase
transition.

In summary, we presented and illustrated a simple promising tool to extract
from the magnetic field dependence of the conductivity at $T_{c}$ the
dynamical critical exponent $z$, the thickness $d$ of thin superconducting
films and interfaces, and the limiting length $\widehat{L}$, giving rise to
rounded BKT- and QSI transitions even in zero field. In fact, in the quantum
case is the divergence of the zero temperature correlation length $\xi
\left( T=0\right) =\overline{\xi }_{0}\delta ^{-\overline{\nu }}$ prevented
because it cannot beyond $\widehat{L}$ and with that is the attainable
tuning regime bounded by $\delta >\left( \overline{\xi }_{0}/\widehat{L}\right) ^{1/\nu }$.

The author is grateful to H. Aubin and C. A. Marrache-Kikuchi for providing
the Nb$_{0.15}$Si$_{0.85}$ resistance data and N. Reyren, S. Gariglio, A. D.
Caviglia, D. Jaccard, and J.-M. Triscone for providing the LaAlO$_{3}$/SrTiO$_{3}$\ interface data. I also thank J.-M. Triscone, N. Reyren, S. Gariglio, A. D. Caviglia, and S. Weyeneth for fruitful conversations.

\end{document}